\documentclass{article}
\usepackage{epsf,amssymb}
\font\grb=eurb10
\newcommand{\bphi}{\mbox{\grb\char'047}\,}
\newcommand{\bomega}{\mbox{\grb\char'041}\,}
\newcommand{\bfi}{\mbox{\grb\char'036}\,}

\def\vpint{\mathop{\int\hskip-0.8pc\mbox{\it /}}}
\newcommand{\rightnote}{{\it \jobname : submitted to
{\bf World Scientific} on September 30, 1999}}%

\begin{document}

\title{Monodromy Transform Approach to Solution of Some
Field Equations in General Relativity and String Theory}
\author{G. A. Alekseev\\
{\it Steklov Mathematical Institute of the Russian
Ac. Sci.,}\\{\it Gubkina str. 8, 117966, GSP -
1, Moscow, Russia}\\
{\it E-mail: G.A.Alekseev@mi.ras.ru}}
\date{September 30, 1999}
\maketitle
\begin{abstract}
A monodromy transform approach, presented in this communication,
provides a general base for solution of space-time symmetry
reductions of Einstein equations in all known integrable cases, which
include vacuum, electrovacuum, massless Weyl spinor field and stiff
fluid, as well as some string theory induced gravity models. There
were found a special finite set of functional parameters  which are
defined as the set of monodromy data for the fundamental solution of
associated spectral problem. These monodromy data consist of the
functions of the spectral parameter only. Similarly to the scattering
data in the inverse scattering transform, the monodromy data can be
used for characterization of any local solution of the field
equations.  A "direct" and "inverse" problems of such monodromy data
transform admit unambiguous solutions. For the linear singular
integral equation with a scalar (i.e. non-matrix) kernel, which
solves the inverse problem of this monodromy transform, an equivalent
regularization -- a Fredholm linear integral equation of the second
kind is constrcuted in several convenient forms. The existence and
uniqueness of the local solution for arbitrary choice of the
monodromy data can be proved using a simple iterative method. This
solution is effectively constructed in terms of homogeneously
convergent functional series.
\end{abstract}

\section{Introduction}
The fundamental nature of Einstein equations as well as beautiful
discovery of the existence of a large class of two-dimensional
completely integrable systems made very natural various
expectations and conjectures of the integrability of the Einstein
equations, at least for the space-times with an
Abelian two-dimensional isometry group, when the reduced dynamical
equations are effectively two-dimensional. First of all, this
concerned the Einstein equations for gravitational fields in vacuum,
which integrability was conjectured and even motivated partially
long ago (see the papers of Geroch, Maison)~\footnote{The lack of
space urges the author to avoid a detail citation and to refer the
reader to the references in a few papers cited below, but mainly --
to a large and very useful F.J.Ernst's collection of
related references and abstracts, accessible throw {\sf
http://pages.slic.com/gravity}, as well as to gr-qc and hep-th data
bases.}. However, the actual discovery of very rich internal
structure of these equations and development of effective methods for
the construction of infinitely large classes of their solutions
actually have been started more then twenty years ago. It is
necessary to mention here a variety of more or less general and well
known now methods and results, such as Belinskii and Zakharov
formulation of the inverse scattering method and their construction
of vacuum $N$-soliton solutions, the constructions of B\"acklund
transformations of Harrison and of Neugebauer, the infinite
dimensional algebra of internal symmetries of stationary axisymmetric
electrovacuum Einstein - Maxwell equations, found by Kinnersley and
Chitre and "exponentiation" of some of these symmetries made by
Hoenselaers, Kinnersley and Xanthopoulos. Later it was shown, that
besides the vacuum case, the integrability properties are possessed
by two-dimensional space - time symmetry reductions of Einstein
equations in the presence of the massless matter fields -- the
electromagnetic fields (Kinnersley and Chitre, Hauser and Ernst,
GA), or/and Weyl massless two-component spinor (neutrino) field
(GA), or/and minimally coupled scalar field, or/and stiff fluid with
$p=\varepsilon$ (Belinskii), or electromagnetic field with dilaton
(Belinski and Ruffini), as well as of some string theory induced
gravity models with axion, dilaton and electromagnetic fields (e.g.,
Bakas, Sen, Gal'tsov and Kechkin).

In this communication we present a sketch of general approach to the
analysis and solution of all mentioned above integrable space - time
symmetry reductions of Einstein equations (of both, hyperbolical and
elliptical types). This approach, called a "monodromy transform
approach", is based on and it develops the results of the author's
papers~\cite{GA:1980a}${}^-$\cite{GA:1988}. It leads a) to a
definition in the most general context of a convenient set of
functional parameters -- "monodromy data", which analytical
properties on the spectral plane are closely related to various
physical and geometrical properties of solutions, and  b) to a
construction of pure linear integral equations (of Cauchy and then,
of Fredholm types) equivalent to the original reduced field equations
and admitting a construction of their general local solutions in
terms of homogeneously converging functional
series~\footnote{In~\cite{GA:1983,GA:1985,GA:1988}
many intermediate statements were argued very briefly and for the
analytical case only. However, all these considerations are valid
also for a larger class of solutions with very low order  of
differentiability (namely, $C^3$ for the metric components). The
corresponding rigorous proof can be found in the
recently published paper of Hauser and Ernst (I.~Hauser, F.J.~Ernst,
gr-qc/9903104), where many closely related statements were proved in
fullest detail, but in a different context of the analysis of the
group structure of the solution space of the Ernst equations and
characteristic initial value problem.}. Various applications of this
approach to a classification of solutions, exact linearization of
various boundary value problems and to explicit construction of new
classes of exact solutions are expected to be considered elsewhere.

\section{Generalized Ernst equations}

The space - time symmetry ansatz of existence of an Abelian two -
dimensional space - time isometry group, provided all field
components and potentials also possess this symmetry, provides
a reduction of all mentioned above cases or eventually, of Einstein
- Maxwell - Weyl equations to generalized form of the Ernst
equations, except for axion - dilaton gravity, which leads
(as it is already known) to a matrix analog of these equations.
In the differential form notation the reduced Einstein
- Maxwell - Weyl equations can be written as~\footnote{These
equations follow immediately from generalized
Kinnersley equations derived in~\cite{GA:1983}.}
\begin{equation}\label{ErnstEquations}
\left\{\begin{array}{l} d\,{}^{\ast}d{\cal
E}+\displaystyle{d(\alpha+i\delta)\over\alpha}\,{}^{\ast}d{\cal
E} -\displaystyle{(d {\cal E}+2\overline{\Phi} d\Phi) \over
\mbox{Re\,}{\cal E}+\Phi \overline{\Phi}}\,{}^{\ast}d{\cal E}
=0\\[1em]
d\,{}^{\ast}d\Phi+\displaystyle{d(\alpha+i\delta)\over\alpha}\,
{}^{\ast}d\Phi -\displaystyle{(d {\cal E}+2\overline{\Phi} d\Phi)
\over \mbox{Re\,}{\cal E}+\Phi
\overline{\Phi}}\,{}^{\ast}d\Phi=0\\[1.2em]
d\,{}^{\star}d\alpha=0,\quad\qquad d\beta \equiv-\epsilon\,
{}^{\star}d\alpha\\[1em] d\,{}^{\star}d\gamma=0,\quad\qquad
d\delta \equiv {}^{\star}d\gamma.
\end{array}\right.
\end{equation}
where ${\cal E}(x^1,x^2)$ and $\Phi(x^1,x^2)$ are  complex scalar
Ernst potentials; "${}^{\ast}$" is a Hodge star operator,
such that $d\,{}^{\ast}d$ is the two-dimensional d'Alambert or
Laplace operator in the hyperbolical ($\epsilon=1$) or elliptical
($\epsilon=-1$) case respectively, defined on the orbit space
$(x^1,x^2)$.  The real functions $\alpha(x^1,x^2)$ -- a measure of
area on the orbits and $\gamma(x^1,x^2)$ -- a potential for neutrino
current vector, are arbitrary "harmonical" functions,  provided
$d\alpha\wedge {}^{\ast}d\alpha\ne 0$.  These functions determine two
other auxiliary real functions -- their "harmonical" conjugates
$\beta(x^1,x^2)$ and $\delta(x^1,x^2)$.

\section{Equivalent "spectral" $N\times N$ - matrix
problem}
For each of the integrable reductions of Einstein equations
considered above we use similar associated complex $N\times N$ -
matrix problems ($N=2$ for vacuum fields, $N=3$ for the models with
electromagnetic and Weyl spinor fields and $N=4$ for string theory
induced gravity models with axion, dilaton and electromagnetic
fields) for the four unknown matrix functions
\begin{equation}\label{Matrices}
{\bf U}(\xi,\eta),\,\, {\bf V}(\xi,\eta),\,\,{\bf
\Psi}(\xi,\eta,w),\,\, {\bf W}(\xi,\eta,w)
\end{equation}
which should satisfy two groups of conditions. The first
one is a deformation problem for a linear system
with given (case dependent) structures of canonical forms
of coefficients and normalization at some reference point
$(\xi_0,\eta_0)$:
\begin{equation}\label{Linsys} \begin{array}{lccl}
\left\{\begin{array}{l}
2 i (w-\xi)\partial_\xi {\bf \Psi}={\bf U}(\xi,\eta)  {\bf
\Psi}\\[1ex]
2 i (w-\eta)\partial_\eta {\bf \Psi}={\bf V} (\xi,\eta)
{\bf \Psi}\end{array}\right.
&\left.\vphantom{\begin{array}{l}
2 i (w-\xi)\partial_\xi {\bf \Psi}={\bf U}(\xi,\eta)  {\bf
\Psi}\\[1ex]
2 i (w-\eta)\partial_\eta {\bf \Psi}={\bf V} (\xi,\eta)
{\bf \Psi}\end{array}}\hskip1ex\right\Vert\hskip1ex&
\left.\begin{array}{l}
({\bf U})_{can}={\bf U}_{(0)}\\[1ex]
({\bf V})_{can}={\bf V}_{(0)}\end{array}
\hskip1ex\right\Vert\hskip1ex& {\bf
\Psi}(\xi_0,\eta_0,w)={\bf I} \end{array}
\end{equation}
The second group of conditions implies the existence for
(\ref{Linsys}) of a Hermitian integral of certain structure with case
dependent constant matrix ${\bf\Omega}$:
\begin{equation}\label{Wequations} \begin{array}{rcl}
\left\{\begin{array}{l}
{\bf \Psi}^\dagger {\bf W} {\bf \Psi} = {\bf W}_0(w)\\[1ex]
{\bf W}_0^\dagger (w)={\bf W}_0 (w)\end{array}\right.&\left.
\vphantom{\begin{array}{l}
{\bf \Psi}^\dagger\cdot {\bf W}\cdot{\bf \Psi} = {\bf W}_0(w)\\[1ex]
{\bf W}_0^\dagger (w)={\bf W}_0 (w)\end{array}}
\hskip3ex\right\Vert\hskip3ex &
\displaystyle{\partial {\bf W}\over\partial w} = 4 i {\bf \Omega}
\end{array}
\end{equation}
where $w$ is complex ("spectral") parameter and $\xi$, $\eta$ are
geometrically defined space-time coordinates:  $\xi=\beta+ j\alpha$,
$\eta=\beta-j\alpha$ with $j=1$ for $\epsilon=1$ and $j=i$ for
$\epsilon=-1$. Thus, for the hyperbolic case ($\epsilon=1$) the
coordinates $(\xi,\eta)$ are two real light cone coordinates, while
for the elliptical case ($\epsilon=-1$) these coordinates are complex
conjugated to each other.  The canonical forms of ${\bf U}$ and ${\bf
V}$ matrices (up to a permutation of diagonal elements) are
$$\left.\begin{array}{l}
{\bf U}_{(0)}^{N=2}=\mbox{diag}\,(i,0)\\[1ex]
{\bf V}_{(0)}^{N=2}=\mbox{diag}\,(i,0)
\end{array}\hskip0.2ex\right\Vert\hskip0.2ex
\left.\begin{array}{l}
{\bf U}_{(0)}^{N=3}=\mbox{diag}\,(i+a(\xi),0,0)\\[1ex]
{\bf V}_{(0)}^{N=3}=\mbox{diag}\,(i+b(\eta),0,0)\end{array}
\hskip0.2ex\right\Vert\hskip0.2ex
\begin{array}{l}
{\bf U}_{(0)}^{N=4}=\mbox{diag}\,(i,i,0,0)\\
{\bf V}_{(0)}^{N=4}=\mbox{diag}\,(i,i,0,0) \end{array}
$$
where $a(\xi)=2\partial_\xi\gamma$, $b(\eta)=2\partial_\eta\gamma$ and
a spinor field potential $\gamma$ is an arbitrary
real solution of
$\partial_\xi\partial_\eta\gamma=0$, provided, for $\epsilon=-1$,
$\vert\mbox{Im}\, a(\xi)\vert<1$
and $\vert\mbox{Im}\,b(\eta)\vert<1$ at least for $\xi$, $\eta$
close enough to $\xi_0$, $\eta_0$.
The matrices ${\bf \Omega}$ are constant:
$${\bf \Omega}^{N=2}=\left(\hskip-1.5ex\begin{array}{rr}
0&1\\-1&0
\end{array}\right)
\hskip1ex\left\Vert\hskip1ex
{\bf \Omega}^{N=3}=\left(\hskip-1.5ex\begin{array}{rrr}
0&1&0\\-1&0&0\\0&0&0 \end{array}\right)
\hskip1ex\right\Vert\hskip1ex
{\bf\Omega}^{N=4}=\left(\hskip-1.5ex\begin{array}{rrrr}
0&0&1&0\\0&0&0&1\\-1&0&0&0\\0&-1&0&0\end{array}\right)$$

For any solution of (\ref{ErnstEquations}) the matrices
(\ref{Matrices}) can be calculated explicitly and for any solution
(\ref{Matrices}) of (\ref{Linsys}), (\ref{Wequations}) the solution
of the (generalized) Ernst equations can be easily calculated
using the identifications
$\partial_\xi{\cal E}=-{\bf U}_1{}^2$, $\partial_\eta{\cal E}=-{\bf
V}_1{}^2$ and $\partial_\xi\Phi={\bf U}_1{}^3$,
$\partial_\eta\Phi={\bf V}_1{}^3$. Besides that,
(\ref{Matrices}) -- (\ref{Wequations}) imply ${\bf W}=4
i(w-\beta){\bf \Omega}+{\bf G}(\xi,\eta)$, where the components of
${\bf G}(\xi,\eta)$ are algebraically related to the
components of metric and electromagnetic
potential~\cite{GA:1980a}${}^-$\cite{GA:1983}.

\section{Direct problem of the monodromy
transform and definition of the monodromy data}
\begin{figure}[h]
\begin{center} \epsfxsize=.97\textwidth \epsfbox{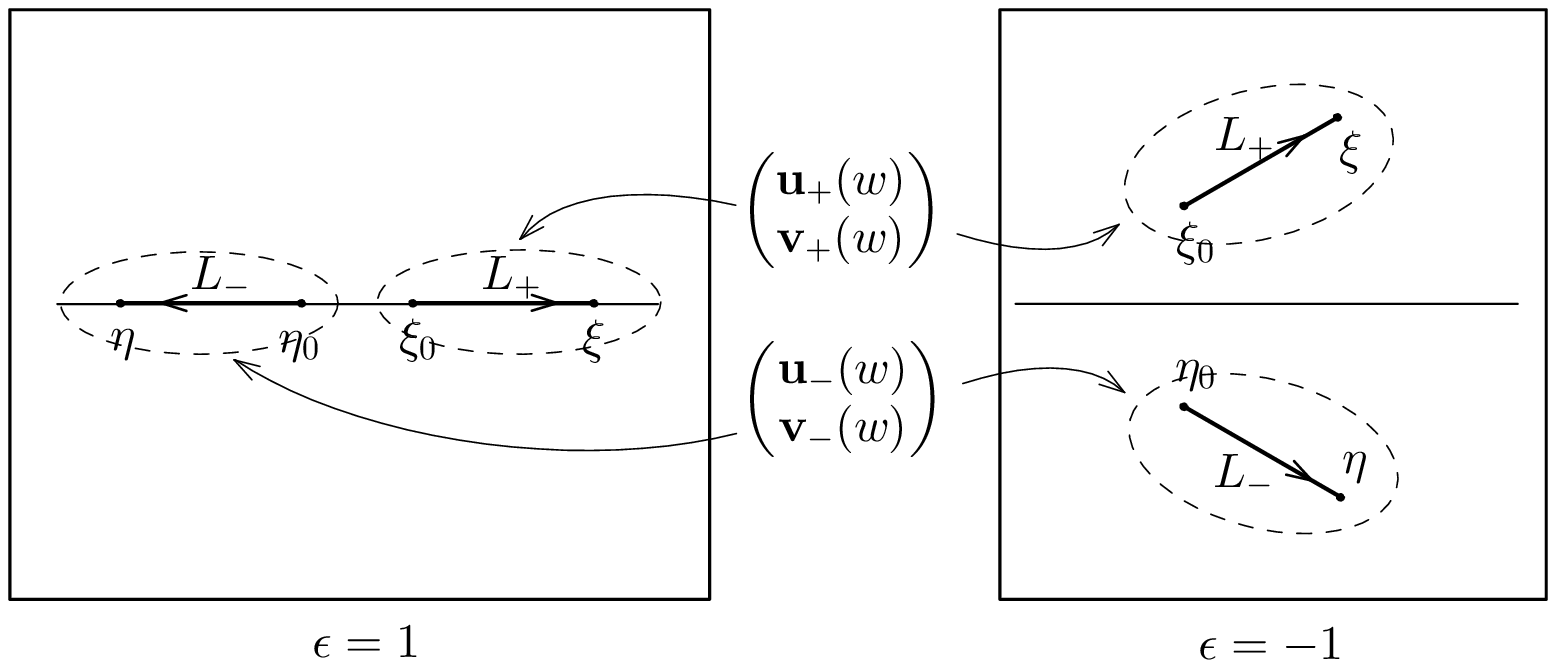}
\caption{The structures of the cut $L=L_+\cup L_-$ on the spectral
plane $w$  and
domains of holomorphicity of the monodromy data functions ${\bf
u}_\pm(w)$ and ${\bf v}_\pm (w)$ in the hyperbolic ($\epsilon=1$) and
the elliptic ($\epsilon=-1$) cases.  \label{fig:wplane}}
\end{center}
\end{figure}

Let us consider at first the linear system (\ref{Linsys}).
It can be shown~\cite{GA:1988}, that any its solution ${\bf
\Psi}(\xi,\eta,w)$ is holomorphic on the spectral plane outside a cut
$L=L_+\cup L_-$, which structure is shown on the
Figure \ref{fig:wplane}.  Four endpoints of this cut
are the branchpoints of ${\bf \Psi}$.  The local behaviour of ${\bf
\Psi}$ at the cuts $L_+$ and $L_-$ is characterized by
monodromy matrices $T_+(w)$ and $T_-(w)$, which describe the
transformations of ${\bf \Psi}$ along the paths, surrounding
the branchpoints on  $L_+$ and $L_-$:
\begin{equation}\label{Tmatrices}
{\bf \Psi}\stackrel {{\bf
T}_\pm}\longrightarrow \widetilde{\bf\Psi}={\bf\Psi}\cdot {\bf
T}_\pm(w),\qquad {\bf T}_\pm(w)={\bf I}-(1+e^{-2 i
[\sigma]_\pm})\displaystyle{{\bf l}_\pm(w)\otimes{\bf k}_\pm(w)\over
({\bf l}_\pm(w)\cdot{\bf k}_\pm(w))}
\end{equation}
where $2 i [\sigma]_+=\pi a(w)$ and
$2 i [\sigma]_-=\pi b(w)$. If a spinor field vanishes, i.e. for
$a(w)=b(w)=0$, the branchpoints at the ends of $L_+$ and $L_-$ are
algebraic branchpoints of the orders $\frac12$ or $-\frac12$ and
therefore, we have ${\bf T}_\pm^2(w)\equiv{\bf I}$.
The structure (\ref{Tmatrices}) of monodromy matrices $T_\pm(w)$
allows to associate with any fundamental solution ${\bf \Psi}$ four
complex vector functions ${\bf k}_\pm(w)$, ${\bf l}_\pm(w)$, defined
(due to a homogeneity of these expressions) in a projective sense and
depending upon the spectral parameter only:
\begin{equation}\label{Mdata}
{\bf k}_\pm(w)=(1,{\bf u}_\pm(w),{\bf v}_\pm(w)),\qquad
{\bf l}_\pm(w)=(1,{\bf p}_\pm(w),{\bf q}_\pm(w)),\qquad
\end{equation}
Following~\cite{GA:1985,GA:1988} we can find that
(\ref{Wequations}) are equivalent to some constraint on the
monodromy data (\ref{Mdata}), which unambiguously relates the
components of the vectors ${\bf l}_\pm$ and ${\bf k}_\pm$ and hence,
the functions ${\bf u}_\pm(w)$ and ${\bf v}_\pm(w)$ can represent a
complete set of the monodromy data for the entire problem
(\ref{Linsys}) - (\ref{Wequations}).
In the case of axion - dilaton gravity instead of vectors
(\ref{Mdata}) we have $2\times 4$ - matrices of the
structure ${\bf k}_\pm(w)=({\bf I},{\bf u}_\pm(w))$, where ${\bf I}$
is a $2\times 2$ unit matrix and ${\bf u}_\pm(w)$ are arbitrary
$2\times 2$ - matrix functions. These monodromy
data can be associated with any local solution of the reduced field
equations and therefore, this construction solves the direct problem
of our monodromy transform.

\section{Inverse problem of the monodromy
transform and equivalent singular integral
equations}
Simple arguments, similar to once used
in~\cite{GA:1985,GA:1988}, show that
certain components of local algebraic structure of
${\bf \Psi}$ on $L_\pm$ -- the components of two complex
vectors $\bphi_\pm(\xi,\eta,w)$ should satisfy the similar sets of
linear singular integral equations.  (In the case of gravity with
axion, dilaton and electromagnetic fields $\bphi_\pm(\xi,\eta,w)$ are
$2\times 4$ - matrices.) Omitting farther the suffices $\pm$ and
keeping in mind that, for example, ${\bf k}(\tau)\equiv{\bf
k}_+(\tau)$ for $\tau\in L_+$ and ${\bf k}(\tau)\equiv{\bf
k}_-(\tau)$ for $\tau\in L_-$, we can write these integral equations
in the form \begin{equation}\label{Line}
\nu(\xi,\eta,\tau)\bphi(\xi,\eta,\tau)+\displaystyle{{1\over\pi
i}\vpint\limits_L}\,{[\,\lambda e^{i\sigma}\,]_\zeta
\over\zeta-\tau}\,{\cal H}(\tau,\zeta)\,
{\bphi}(\xi,\eta,\zeta)\,d\,\zeta =-{\bf k}(\tau) \end{equation}
where a Cauchy principal value integral is used, and the
coefficients are
$$\nu(\xi,\eta,\tau)=-\{\lambda e^{i\sigma}\}_\tau{\cal
H}(\tau,\tau),\qquad
{\cal H}(\tau,\zeta)=({\bf k}(\tau)\cdot{\bf l}(\zeta);$$
$[\ldots]_\zeta$ and $\{\ldots\}_\zeta$ are a "jump"
and a "continuous part" of functions at the point $\zeta\in L$.
The functions $\lambda(\xi,\eta,w)=\sqrt{(w-\xi)(w-\eta)
/(w-\xi_0)(w-\eta_0)}$ and
$2\sigma(\xi,\eta,w)=\int_{L_+}a(\zeta)/(w-\zeta)d\zeta+
\int_{L_-} b(\zeta)/(w-\zeta)d\zeta$.

Thus, the integral equation (\ref{Line}) is determined completely in
terms of functions (\ref{Mdata}). Besides that, the Ernst potentials
and all of the field components can be calculated as path integrals,
which are also determined in terms of monodromy data and the
corresponding solution of (\ref{Line})~\cite{GA:1985,GA:1988}.
Therefore, the solution of (\ref{Line}) solves the inverse problem
of our monodromy transform.

\section{Equivalent Fredholm equation: the existence and uniqueness
of local solutions for arbitrary chosen monodromy
data}

In accordance with the well known theory of linear singular integral
equations, the equation (\ref{Line}) within the class of solutions
regular on the cut, possesses an important property, that
the index of its characteristic part is equal to zero for arbitrary
chosen monodromy data functions. This means that this equation
admits various equivalent regularizations. We present two
equivalent forms of the corresponding (quasi-) Fredholm equations
which are left and right regularizations of (\ref{Line})
(the dependence upon $\xi$, $\eta$ is not shown here explicitly):
\begin{equation}\label{Flines}
\bfi(\tau)+\int\limits_L{\cal
F}(\zeta,\tau)\bfi(\zeta)\,d\,\zeta={\bf h}(\tau),\qquad
\bomega(\tau)+\int\limits_L{\cal
G}(\zeta,\tau)\bomega(\zeta)\,d\,\zeta={\bf k}(\tau) \end{equation}
where $\bfi(\tau)=-{\cal H}(\tau,\tau)\bphi(\tau)$ and the following
relations take place
$$\bfi(\tau)=-\displaystyle{1\over
B(\tau)Z(\tau)} {\cal R}_\tau \left[B(\tau)\bomega(\tau)\right],\qquad
{\bf h}(\tau)=-\displaystyle{1\over
B(\tau)Z(\tau)} {\cal R}_\tau \left[B(\tau){\bf k}(\tau)\right]$$
The kernals ${\cal F}(\zeta,\tau)$ and ${\cal G}(\zeta,\tau)$
are determined by the expressions
$$\left.\begin{array}{l}
{\cal F}(\zeta,\tau)=\displaystyle{B(\zeta)Z(\zeta)\over
B(\tau)Z(\tau)}{\cal R}_\tau \left[B(\tau){\cal
S}(\tau,\zeta)\right]\\[2ex] {\cal
G}(\zeta,\tau)=B(\zeta)\widetilde{\cal R}_\zeta \left[{\cal
S}(\tau,\zeta)\right]
\end{array}\hskip2ex\right\Vert\hskip2ex{\cal
S}(\tau,\zeta)=\displaystyle{{\cal H}(\tau,\zeta)-{\cal
H}(\zeta,\zeta)\over i\pi {\cal H}(\zeta,\zeta)(\zeta-\tau)}$$
where the operators ${\cal R}_\tau$, $\widetilde{\cal R}_\tau$ and auxiliary
functions possess the expressions
$$\left.\begin{array}{l}
{\cal R}_\tau
\left[f(\tau)\right]=A(\tau)
f(\tau)-B(\tau)Z(\tau)\displaystyle{{1\over
i\pi}\vpint\limits_L {f(\zeta)\,d\,\zeta\over
Z(\zeta)(\zeta-\tau)}}\\[1ex]
\widetilde{\cal R}_\tau
\left[f(\tau)\right]=A(\tau) f(\tau)+\displaystyle{{1\over
Z(\tau)}{1\over i\pi}\vpint\limits_L\displaystyle{B(\zeta)
Z(\zeta)\over (\zeta-\tau)}f(\zeta)\,d\,\zeta}
\end{array}\hskip2ex\right\Vert\hskip1ex
\begin{array}{l}
A(\tau)=\sin [\sigma]_\tau\\[1ex]
B(\tau)=i \cos [\sigma]_\tau\\[1ex]
Z(\tau)=i[\,\lambda\,]_\tau e^{i\{\sigma\}_\tau}
\end{array}$$
(We note here, that for electrovacuum ($\sigma\equiv0)$ we have
$A(\tau)=0$, $B(\tau)=i$.)

The local solution of each of the equations (\ref{Flines}) for any
given set of monodromy data can be constructed by the known
iterative method. In particular,
\begin{equation}\label{Series}
 \begin{array}{lcl}
\bfi(\tau)=\bfi_0(\tau)+\sum\limits_{n=1}^\infty
\left(\bfi_n(\tau)-\bfi_{n-1}(\tau)\right),\\[2ex]
\bfi_0(\tau)={\bf h}(\tau),\qquad
\bfi_n(\tau)={\bf h}(\tau)-\displaystyle{\int\limits_L}{\cal
 F}(\tau,\zeta)\bfi_{n-1}(\zeta)\,d\,\zeta \end{array}
\end{equation}
For local solutions, when the coordinates $\xi$ and $\eta$
are close enough to their initial values $(\xi_0,\eta_0)$, it is
easy to prove a homogeneous convergence of the series (\ref{Series})
and therefore, the existence as well as the uniqueness of the
solution.

\section*{Acknowledgments}
This work was supported in part by the Russian Foundation for Basic
Research Grants 99-01-01150, 99-02-18415.

\end{document}